\documentstyle[12pt,psfig,preprint]{aastex}

\def\simgt{\lower.5ex\hbox{$\; \buildrel > \over \sim \;$}}
\def\simlt{\lower.5ex\hbox{$\; \buildrel < \over \sim \;$}}

\def\amin{\ifmmode^{\prime}\else$^{\prime}$\fi}
\def\asec{\ifmmode^{\prime\prime}\else$^{\prime\prime}$\fi}
\def\nhunits{\times 10^{22}\ {\rm cm^{-2}}}

\begin{document}

\title{An X-ray Image of the Composite SNR G16.7+0.1}

\author{David J. Helfand$^1$, Marcel A. Ag\"ueros$^2$, E. V. Gotthelf$^1$}
\altaffiltext{1}{Columbia Astrophysics Laboratory, Columbia University, 550 West 120$^{th}$ Street, New York, NY 10027, USA; djh@astro.columbia.edu, evg@astro.columbia.edu}
\altaffiltext{2}{Department of Astronomy, University of Washington, Box 351580, Seattle, WA 98195, USA; agueros@astro.washington.edu}

\begin{abstract}
 
We have observed the Galactic supernova remnant G16.7+0.1 for 13 ks
using the EPIC cameras aboard the {\sl XMM-Newton} X-ray Observatory, producing
the first X-ray image of the remnant. This composite radio remnant has a core
radio flux density of only 100~mJy, making it one of the faintest radio
synchrotron nebulae yet detected, although the core-to-shell flux ratio at 6 cm 
is typical of the growing class of composite remnants. Our image is seriously
contaminated by bright arcs produced by singly reflected X-rays from the 
X-ray binary GX17+2 which lies just outside the field of view, providing 
an interesting data analysis challenge. Nonetheless, the remnant's 
synchrotron core is clearly detected. We report on the spectrum and 
intensity of the core emission as well as on our search for emission from 
the thermal shell, and describe the constraints these observations provide 
on the remnant's distance, age, and central pulsar properties.
\end{abstract}


\section {Introduction}

Recently, imaging above 3.5 keV has led to a flurry of discoveries of X-ray synchrotron nebulae and, in some cases, to the pulsars which power them. Some
nebulae, in common with the prototypical Crab Nebula, show no evidence of a surrounding shell created by the supernova explosion. But the majority of such objects are ``composites" showing a flat-spectrum, polarized radio and power-law X-ray synchrotron nebula, presumably powered by a central pulsar, encircled by a typical
supernova remnant (SNR) shell. X-ray observations of both Crab-like and composite remnants provide crucial information on the poorly known initial distribution of pulsar magnetic field strengths and spin periods, as well as on the ages and dynamics of the supernovae which created them. 

G16.7+0.1 is a classic composite remnant with one of the faintest radio core components detected to date (Helfand et al. 1989). It has comparable radio luminosities in the core and shell components; the shell and core radii are $\sim 2$\amin\  and $\sim 1$\amin, respectively. A recently published catalog of
X-ray sources derived from the {\sl ASCA} Galactic Plane Survey lists G16.7+0.1 as a detection (Sugizaki et al. 2001), although the {\sl ASCA} angular
resolution is insufficient to resolve the source. We 
undertook imaging spectroscopy of the source with {\sl XMM-Newton}
in order to separate the shell and core components, determine their 
luminosities, and search for the central pulsar.

In section 2 we describe our observations and the analysis procedures
employed. Section 3 presents the results of this analysis for the remnant's X-ray core, surrounding shell, and presumably associated (but undetected)
pulsar. We conclude in section 4 with a discussion of the implications of these results for G16.7+0.1, and place this source in the context of the
class of composite remnants.

\section{Observations and Analysis Procedures}

G16.7+0.1 was observed on 2001 March $8-9$ for 13 ks with the EPIC camera 
on board the
{\sl XMM-Newton} Observatory (Jansen et al. 2001) (see Figure 1). Data
were obtained from all three cameras which comprise the EPIC (European Photon 
Imaging Camera) instrument. The two MOS cameras (Turner et al. 2001) and the PN
camera (Str\"uder et al. 2001) are CCD arrays sensitive to photons with energies
between 0.1 and 15 keV; the field of view is 30\amin\ in diameter. 
The CCD pixel sizes are 1.1 and 4.1\asec, respectively, while 
the mirror point spread function (PSF) is $\sim6$\asec\ 
full-width half maximum. 
G16.7+0.1 was observed with the medium filter and in full-image mode, so
that the time resolution is 2.5 s for the MOS data and 73 ms for the PN.
We verified the images' astrometric accuracy by optically identifying 
the bright source on the central MOS chip (at {\sl XMM} 
reported RA(J2000) $18^{\rm h} 21^{\rm m} 17^{\rm s}.35$, 
DEC(J2000) $-14^{\rm o} 15^{\amin} 32^{\asec}.6$ for MOS1 and 
RA $18^{\rm h}21^{\rm m} 17^{\rm s}.30$, 
DEC $-14^{\rm o} 15^{\amin} 33^{\asec}.9$ for MOS2). 
This source is coincident with a USNO catalog star; its position agreed 
with the average MOS position to within 0.01 s in Right Ascension
and 3.15\asec\ in Declination, within the astrometric accuracy expected.   

The {\sl XMM} Standard Analysis System (SAS\footnote{version 5.2+, xmmsas-20010728-0329})
allows one to filter the data for good events within the camera's energy range, 
thereby removing most contamination from internal background and cosmic ray particles. In addition, we examined the light curves for each instrument and 
removed intervals corresponding to high background activity. The resulting 
filtering produced 8.8 ks of data for the two MOS cameras, and 5.5 ks for the PN
camera. 

The absorption column density to the remnant is high, suggesting a relatively 
large distance to the source. As a consequence, few source photons are detected
at energies $\leq 1$ keV. Unless otherwise noted, all spectral fitting was done 
over the energy range of $1$ to $\sim 8$ keV, with the upper limit 
being slightly different for each camera. Furthermore, the position of
the remnant on the PN camera was perilously close to the gap between
two chips. These data were therefore used exclusively to verify consistency
with our primary results, obtained from MOS1 and MOS2. 

Bright arcs are easily visible in the data from all three EPIC detectors.
We believe that these are due to single-scatter photons from the nearby 
(but outside the field of view) bright X-ray binary GX17+2. 
In order to characterize the contribution of these arcs to the source 
background (where they are dim but present), we first compared the 
spectra of different arc regions 
to detect possible variations. Finding none, we divided the central MOS 
chip into three polygonal areas. This geometry mimics the elliptical shape 
of the arcs, so that a given 
region contains a full section of any arc present in that part of the chip.

Figure \ref{region} shows the central chip in the filtered MOS1 image. The area 
to the right of the remnant is used to characterize the arcs, and the area 
to the left to characterize the ``normal'' background (the bright USNO
source is excluded from the latter region). These regions are defined 
exclusively on the central chip to avoid having to account for chip-to-chip
variations. The central polygon corresponds to the source background, and
encloses two circles centered on the source; the larger includes
the entire radio remnant (R = 135$^{\asec}$), while the smaller 
zeroes in on the brightest emission region, comparable in size to the 
radio synchrotron core (R = 45$^{\asec}$). The circles are 
centered at RA(2000) $18^{\rm h} 20^{\rm m} 57^{\rm s}.8$, DEC(2000) 
$-14^{\rm o} 20^{\amin} 09^{\asec}.6$, 
consistent with the SNR's apparent geometrical center at 6 cm (Helfand et al.
1989). The areas are obviously not equal and scalings are applied when 
correcting or comparing them.

Figure \ref{model} summarizes the process by which we modeled the arcs'
contribution to the source background. The dotted line corresponds to the 
normal field background; the dashed line is the arcs' spectrum, corrected 
to remove the contribution of this ``standard'' noise. The supernova
remnant is detected against a background which is a weighted
sum of the arc and the normal background contributions. We therefore
compared a number of spectra obtained by summing the two to a spectrum 
directly extracted from the central polygon (without the remnant).   
In Figure \ref{model}, the source background spectrum is the solid curve,
while the dot-dashed curve corresponds to the best-fit, weighted sum of 
normal background and arc spectra (background spectrum + 0.22 arc spectrum). The residuals, obtained by subtracting 
this model from the direct estimate of the source background, are included.
This process was repeated for MOS2, using regions defined similarly on the chip
to enclose the arc and source emission and then weighting by factors appropriate
to the region areas (background spectrum + 0.6 arc spectrum) to produce a
model spectrum. The models obtained for the background allowed us to quantify
and remove the contamination of our observation of G16.7+0.1 by the arcs due 
to GX17+2.

These background models were applied using locally generated RMF and ARF files,
and spectra grouped to contain a minimum of $25\ {\rm counts\ bin^{-1}}$. 
We also used the canned response files, available from the 
{\sl XMM-Newton} website\footnote{http://xmm.vilspa.esa.es/ccf/epic/}, to
check for consistency. All errors are quoted for a 1-$\sigma$ 
confidence range.

\section{Analysis and Results}

\subsection{The X-ray core}
\begin{figure}
\centerline{\psfig{figure=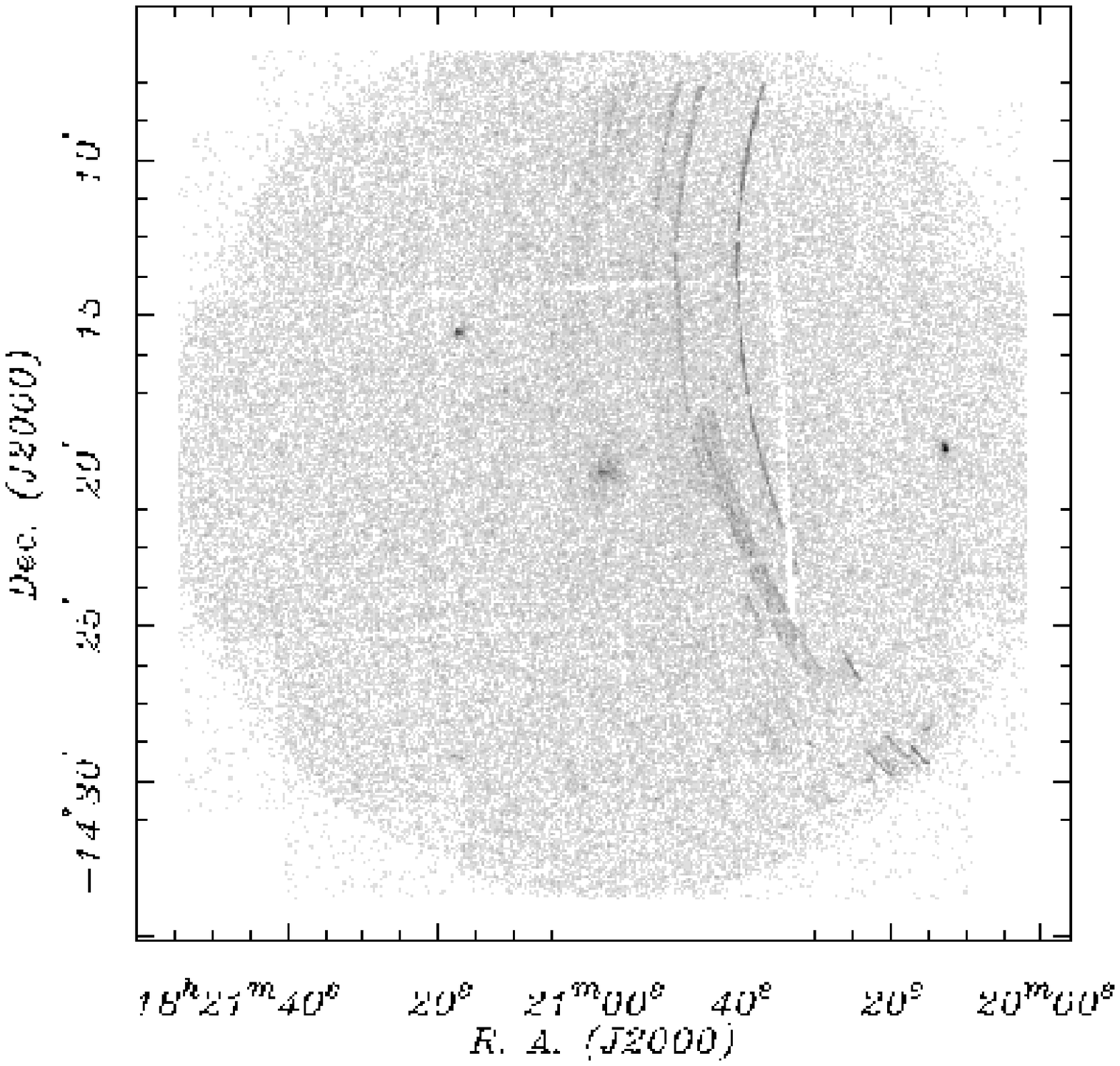,height=3.6in}\psfig{figure=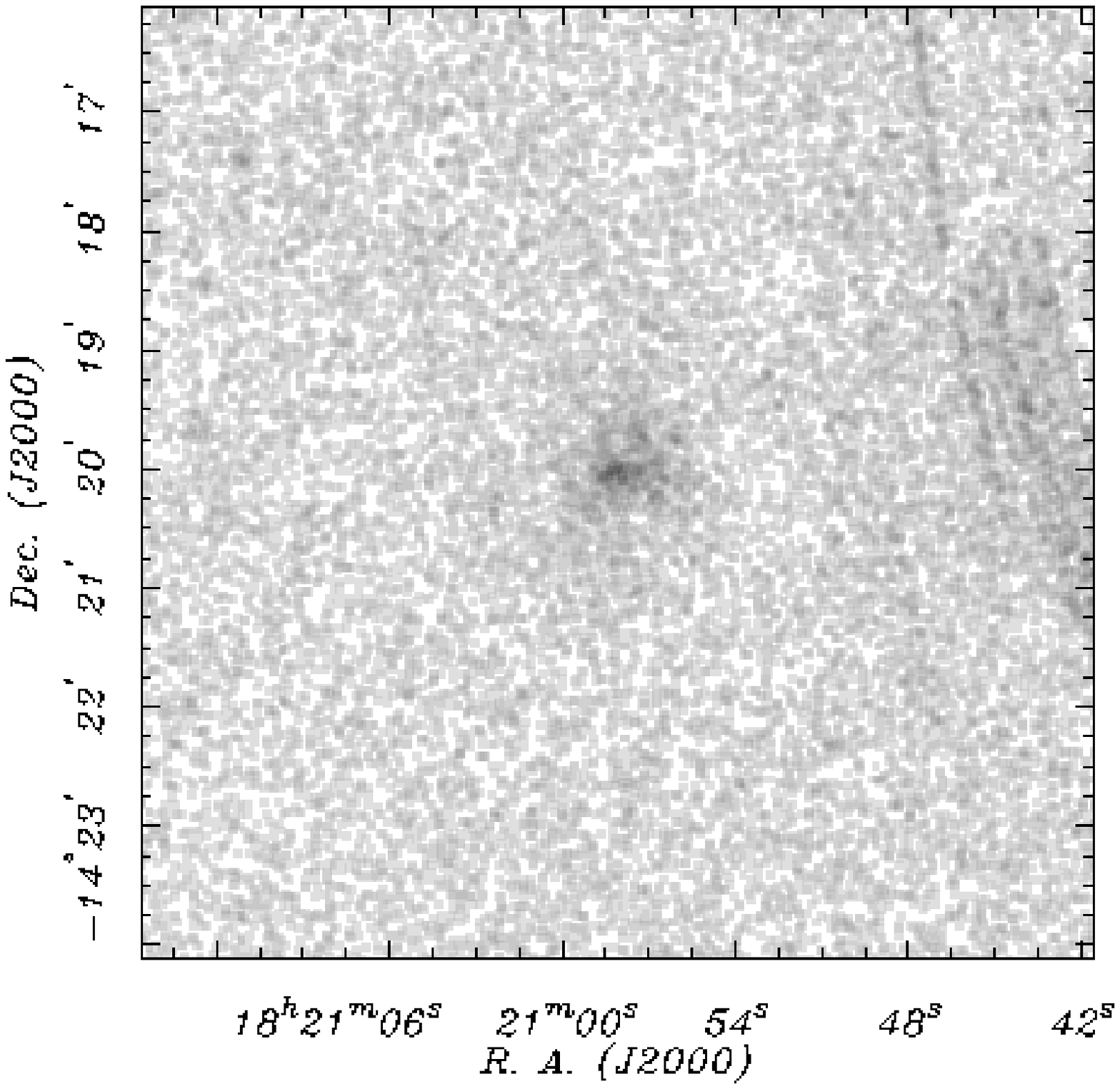,height=3.6in}}
\caption{Full merged MOS1 and MOS2 field of view (left), along with a closeup of G16.7+0.1. The images are displayed with logarithmic intensity.}
\label{picture}
\end{figure}

\begin{figure}
\centerline{\psfig{figure=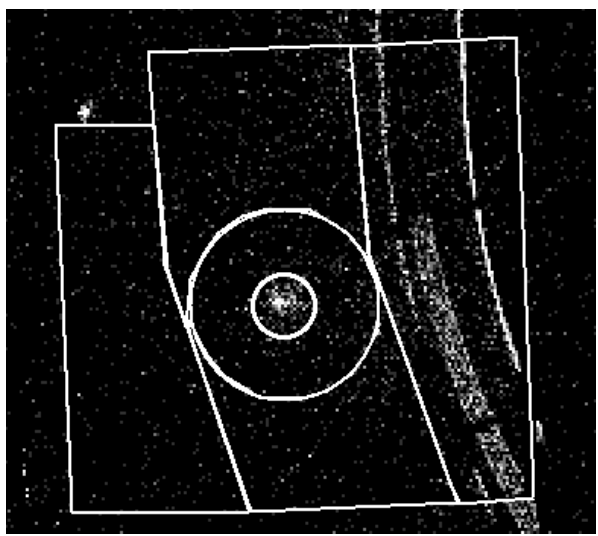,height=2.75in,clip=}}
\caption{Regions used in analyzing the filtered MOS1 data. The outer boundaries correspond to the central chip limits. The large circle, of radius 135 arcseconds, corresponds to the extent of the radio shell; the small circle, of radius 45 arcseconds, encloses the brightest X-ray emitting region. The center of the circles, (RA(2000) $18^{\rm h} 20^{\rm m} 57^{\rm s}.8$, DEC(2000) $-14^{\rm o} 20^{\amin} 09^{\asec}.6$), was derived from radio observations of the source.}
\label{region}
\end{figure}

\begin{figure}
\centerline{\psfig{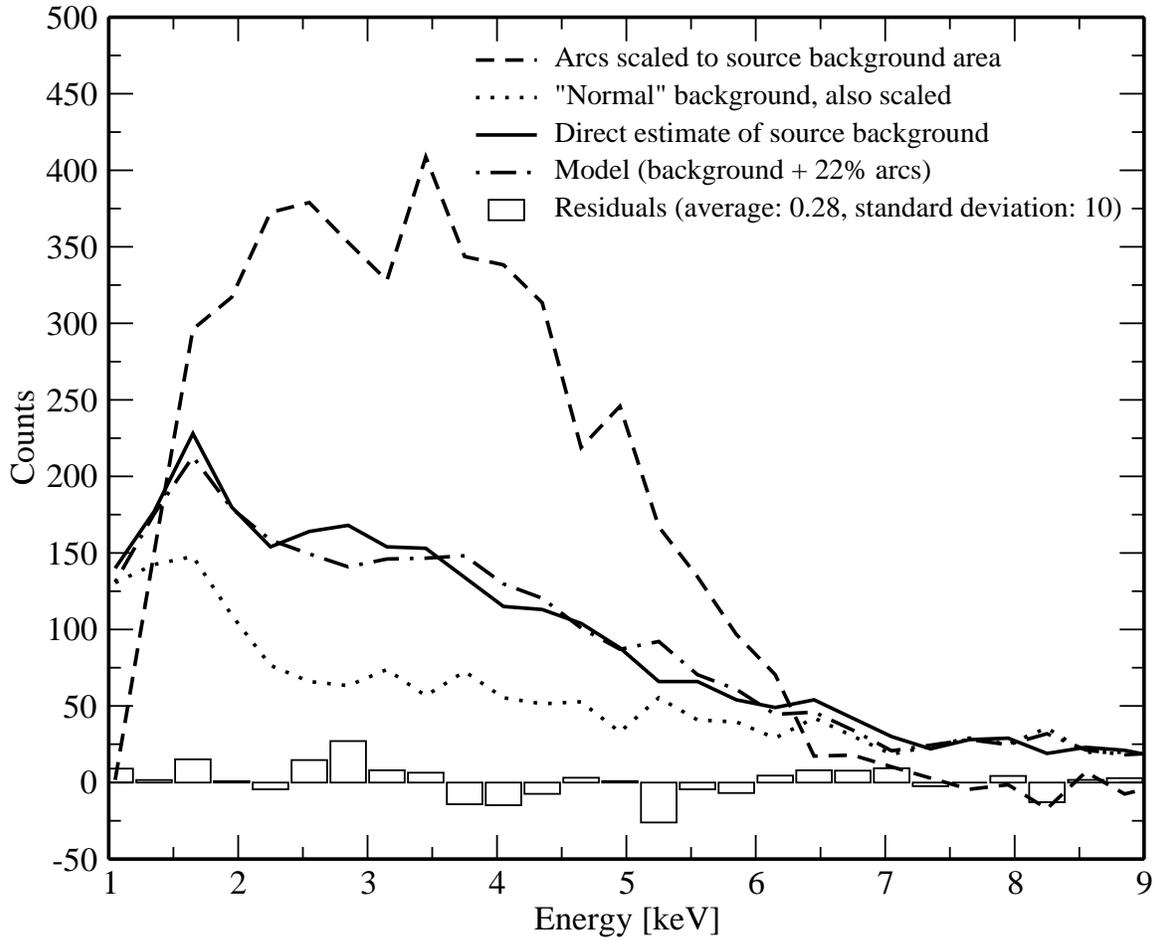}}
\caption{Modeling the contribution of the bright arcs to the source background for MOS1; the source itself has been removed. All the counts have been normalized to the same area.}
\label{model}
\end{figure}

We began by analyzing the bright X-ray core of G16.7+0.1 which we take to 
represent a pulsar wind nebula (PWN; see Figure \ref{picture}).
Spectra were extracted from the MOS images for a circular region of radius
45\asec\ as described above. We used the standard XSPEC spectral analysis 
software to fit the spectra simultaneously, with the energy range included being
slightly different for the two data sets: 1 to 8.4 keV for MOS1 
and 1 to 7.5 keV for MOS2. As noted above, the column density to the
remnant is high, and few photons below 1 keV are detected; the
upper limits are set by the data quality. 

\begin{figure}
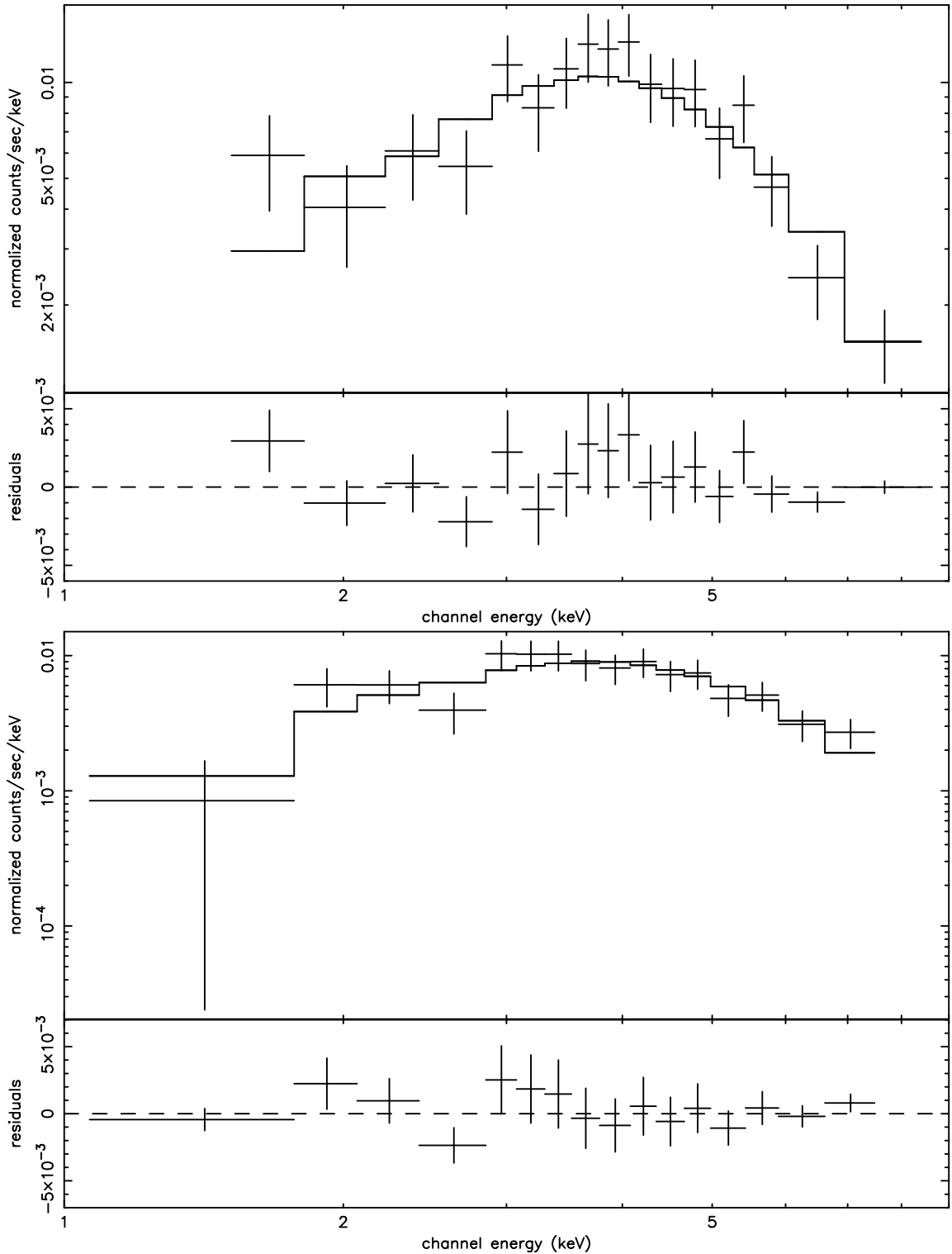

\centerline{\psfig{figure=f4a.eps,width=6in,angle=270}}
\centerline{\psfig{figure=f4b.eps,width=6in,angle=270}}
\caption{Spectral fitting results for the MOS1 (top) and MOS2 (bottom) spectra. The data were fitted simultaneously but are presented separately for clarity. The response files were locally generated, and the models described in the text were used for the background subtraction.}
\label{fit}
\end{figure}

We found that a power-law spectrum modified by interstellar absorption provides
an excellent description of our data for the X-ray core. We find 
a photon index $\Gamma = 1.17 \pm 0.29$ and a column density
$N_H = 4.74 \pm 0.98 \nhunits$, with a reduced $\chi^{2}_\nu = 0.82$ (31 dof)
for a fit using the background models created as described above 
and locally generated RMF and ARF files (see Figure \ref{fit}); the results 
obtained using the canned response files are essentially identical, with
$\chi^{2}_\nu = 0.84$ (31 dof). Fixing $\Gamma$ and
$N_H$ to these values when fitting the extracted PN spectrum (for a 
smaller circular area of R = 28\asec\  chosen to avoid portions of the
source which fell in the interchip gap) produced a $\chi^{2}_\nu = 0.79$ (12 dof) over
the energy range 1 to 10 keV. The nominal power law slope for G16.7+0.1
is slightly flatter than that of any known PWN (see Gotthelf 2003 for a recent 
summary), although within the errors it is consistent with the slopes of the young
pulsars in G11.2-0.3 and G54.1+0.3.

For comparison with these results, we retrieved the {\sl ASCA} data from
the HEASARC\footnote{This research has made use of data obtained from the High 
Energy Astrophysics Science Archive Research Center (HEASARC), provided by 
NASA's Goddard Space Flight Center.} archive and extracted 
the Gas Imaging Spectrometer (GIS) data in the 0.5-8 keV band
within a radius of $6^{\amin}$. We find a background-subtracted count rate
of $0.017\pm 0.001$ counts s$^{-1}$ per GIS. Using the spectral parameters from the MOS
fits, the HEASARC tool PIMMS predicts a 1--8 keV MOS count rate of 0.042 counts
s$^{-1}$, within $4\%$ of the observed value.
Returning to the {\sl XMM} data, we estimated the hardness ratio of the core for
the two cameras (Table 1). The hardness ratio, defined as (N$_{hard}-$ N$_{soft}$)/(N$_{hard}+$ N$_{soft}$), where N is the number of counts, is 
$0.56\pm0.03$, precisely that expected for the spectral parameters derived from our fits.

\begin{table}
\begin{center}
\begin{tabular}{lcccc}
\multicolumn{5}{l}{{\bf Table 1:} Core Hardness}\\
\hline
 & Observation &\multicolumn{2}{c}{Counts (s$^{-1}$)}& Hardness \\
 & Length (s)  & 1 to 3 keV & 3 to 8 keV & Ratio \\
\hline\hline
MOS1 & 8763.5 & 8.18$\pm0.97\times 10^{-3}$ & 3.25$\pm0.19\times 10^{-2}$ & $0.6
0\pm0.04$ \\
MOS2 & 8801.9 & 8.27$\pm0.97\times 10^{-3}$ & 2.68$\pm0.17\times 10^{-2}$ & $0.5
3\pm0.05$\\
\hline
\end{tabular}
\end{center}
\end{table}

The total Galactic HI column density in the direction of G16.7+0.1 is 
$1.56 \nhunits$. Generally, the X-ray-derived value for $N_H$ is two to 
three times greater than the neutral hydrogen column density derived from 21 cm
observations for a given object, owing to contributions from X-ray absorbing
atoms in the molecular and ionized phases of the interstellar medium
(Gorenstein and Tucker 1976). The high
X-ray value we derive is comparable to that found for the composite remnant
Kes 75 (Helfand, Collins, \& Gotthelf 2003) which has a measured distance
of 19 kpc. This suggests that the G16.7+0.1 is at the very least on the 
far side of the Galactic Center, and we adopt D = 10 kpc for our calculations.
The unabsorbed flux for the SNR's 
bright core is $1.9 \times 10^{-12}$ ergs cm$^{-2}$ s$^{-1}$ in the 0.5 
to 10 keV band, corresponding to an X-ray luminosity of $2.3 \times 10^{34} \ \rm D^{2}_{10\  \rm kpc}$ ergs s$^{-1}$.

\subsection{The extended nebula}

In order to compare the outer regions of the X-ray remnant with the known radio
shell component, we attempted to extract spectra for an area equivalent to the remnant shell as defined by the radio data (Helfand et al. 1989), an annulus
with inner and outer radii of $45^{\asec}$ and $135^{\asec}$,
respectively. In all other
respects the analysis was the same as for the core, with the background
models being scaled appropriately for the the area of the annulus. 
 
Unfortunately the small number of counts (roughly 100 for each camera)
from the X-ray shell does not allow us to constrain the nature of this emission
or even to comment on its hardness ratio relative to that of the core. The
matter is further complicated by the likely presence of a dust-scattering
halo from the pulsar wind nebula. As noted by Helfand et al. (2003) in their
analysis of Kes 75, from 10-20\% of the PWN flux in the 1-3 keV band
is scattered into a halo with an angular diameter of $\sim 4^{\amin}$ for
column densities similar to that measured here. From the count rates recorded
in Table 1, and applying the technique outlined in Helfand et al. (2003), it
appears that only $\sim 10\%$ of the photons recorded
in our annulus are attributable to a dust-scattering halo in the soft band,
while at higher energies the scattered fraction declines substantially. Thus,
the remnant does appear to have a net flux associated with the radio shell.
We therefore use PIMMS to determine the flux, over the 0.5 to 10 keV 
range, of a putative Raymond-Smith thermal plasma component (with solar
abundances) responsible 
for emission from this SNR shell. Fixing the column density to our best 
fit value, we obtain the intrinsic luminosities listed in Table 2 for a 
range of possible temperatures (not including any correction for a dust halo). 

\begin{table}
\begin{center}
\begin{tabular}{cc}
\multicolumn{2}{l}{{\bf Table 2:} X-ray Shell Luminosity}\\
\hline
kT    & $L_{Xs}$\\
(keV) & (D$^{2}_{10 \rm \  kpc}$ ergs s$^{-1}$)\\
\hline\hline
0.5   & $2.2 \times 10^{35}$\\
1     & $2.7 \times 10^{34}$\\
1.5   & $1.1 \times 10^{34}$\\
\hline
\end{tabular}
\end{center}
\end{table}

Assuming D = 10 kpc and R = 135\asec\  for the SNR shell gives 
a radius of 6.5 pc, while the R = 45\asec\  core has a radius of 2.2 pc. 
The distance-independent ratio of core-to-shell diameters
is therefore
0.3, consistent with the ratios found from 6 cm observations of other 
composite remnants (Helfand \& Becker 1987). Furthermore, assuming that it is 
in the free-expansion phase with $v_{exp} \approx 3 \times 10^3$ km~s$^{-1}$,
G16.7+0.1 is roughly 2100 years old.

\subsection{A central point source} 

It is interesting to place an upper limit on any contribution from a point
source to the flux detected from the bright X-ray core. 
Ghizzardi (2001) provides the best fitting parameters for in-flight
measurements of the PSF shape\footnote{$r_c=4.11 \pm 0.034$ and $\alpha=1.415 \pm 0.005$ for the King profile and an on-axis source; the mean energy is 1.8 keV for this fit.}.
Comparing the PSF
to the profile of the bright core for the MOS1 and MOS2 data over the
1 to 7 keV range for which the PSF parameters are very nearly constant,
we find that a point source can contribute no more than 37\% of the core's
flux. The consequent upper limit to the source luminosity of
$8.5 \times 10^{33} \ \rm D^{2}_{10\  \rm kpc}$ ergs s$^{-1}$ is 
comparable to that of the young pulsar at the center of G11.2-0.3 
(Torii et al. 1997) and fifty times less
than that from the luminous, young pulsar in Kes 75 (Helfand et al.
2003).
 
\section{Discussion}

The PWN in G16.7+0.1 has a radio luminosity of $0.1 \times 10^{34}$ ergs
s$^{-1}$. Thus we find a value for the core luminosity ratio 
$L_{Xc}/L_{Rc} \sim 20$, slightly above the
mean value of 15 found for sixteen composite remnants compiled by Helfand
(2002). It is similar to that of G11.2-0.3 
($L_{Xc}/L_{Rc}\geq 27$) and CTB80 ($L_{Xc}/L_{Rc}\sim15$), 
and intermediate between the very young and energetic objects such 
as Kes 75 ($L_{Xc}/L_{Rc}\sim140$), 
and older remnants such as Vela ($L_{Xc}/L_{Rc}\sim0.05$). Its X-ray shell
to core luminosity ratio ($\sim 1$) is poorly constrained, but is roughly
similar to that of G11.2-0.3,
as is its shell to core diameter ratio (both are $\sim 4$); for a distance of
10 kpc, the physical diameters of both components are also comparable. Thus,
the X-ray and radio properties of the remnant are consistent with that of
an object with an age of a few thousand years housing a moderately energetic
pulsar.

Seward \& Wang (1988) have outlined a method for obtaining the characteristics
of a pulsar from the PWN luminosity over the 0.2 to 4 keV range.
Our observations of G16.7+0.1 yield a luminosity of $1.1 \times 10^{34}$ ergs
 s$^{-1}$ in this energy range; this produces an estimate that the unseen 
pulsar's $\rm \dot{E}=2.7 \times 10^{36}$ ergs s$^{-1}$. 
Table 3 gives the characteristics of the pulsar derived from this $\rm \dot{E}$ 
as a function of age assuming the pulsar's initial spin period $P_o << P$.

\begin{table}
\begin{center}
\begin{tabular}{cccc}
\multicolumn{4}{l}{{\bf Table 3:} G16.7+0.1 Pulsar Characteristics}\\
\hline
Age & P & $\rm \dot{P}$ & $\rm B_0$ \\
($\times 10^3$ years) & (s) & ($\times 10^{-15}$ s s$^{-1}$) & ($\times 10^{12}$ G)\\
\hline\hline
2.1 & 0.332 & 2498 & 29 \\
5   & 0.215 & 680  & 15 \\
10  & 0.152 & 240  & 6 \\
\hline
\end{tabular}
\end{center}
\end{table}

More recently, van der Swaluw and Wu (2001) have suggested a method of
estimating the pulsar initial spin periods $P_o$ in composite remnants from the 
relative sizes of the PWN core and the remnant shell. For G16.7-0.3, they
infer a value $P_o = 0.043$ s for $\dot{P} = 2.5 \times 10^{-12}$, which yields
the $\dot{E}$ quoted above. Using the well-know relations between the
pulsar spin parameters and magnetic fields, and assuming a braking index
of 3, this initial spin period and an age of 2000 years leads to a current
spin period of 0.33 s and a rather high magnetic field strength of $3 \times 
10^{13}$ G. Such a pulsar would have a spin-down luminosity $\dot {E} = 2.7
\times 10^{36}$ erg s$^{-1}$ (as above), sufficient to drive the observed X-ray
luminosity of the PWN if the efficiency for the conversion of rotational
kinetic energy to X-rays was as low as 0.8\%, significantly below that
of the most efficient X-ray producers (e.g., G29.7-0.3 at 6.5\%; Helfand et al.
2003). For a 5000-year old remnant, the current period and magnetic field 
strength would be $P = 0.17$ s and a more moderate $B = 8 \times 10^{12}$ G.

In conclusion, this first X-ray image of G16.7-0.3 finds it to be consistent
with a classic composite SNR with a central PWN driven by a pulsar with
typical parameters. All that remains to confirm this picture is the detection
of the pulsar itself and the measurement of its spin parameters. A scheduled
{\sl Chandra} observation should provide the detection and point-source
strength, allowing a followup observation with {\sl XMM} to detect the pulse
period.

\begin{acknowledgements}
{\noindent \bf Acknowledgments} --- D.J.H. acknowledges support from grant NAG5-9928. E.V.G. is supported by NASA LTSA grant NAG5-7935. We thank the {\sl XMM} help desks for their assistance and the anonymous referee for a meticulous
review that led to significant improvements in the manuscript.
\end{acknowledgements}

\clearpage

\clearpage

\end{document}